\documentclass[12pt,preprint]{aastex}

\def\mearth{M_\oplus}

\shorttitle{Volatile enrichments and composition of Jupiter}
\shortauthors{Alibert, Mousis \& Benz}

\begin{document}

\title{On the volatile enrichments and composition of Jupiter}

\author{Yann Alibert${}^1$, Olivier Mousis${}^1$ \& Willy Benz${}^1$}

\altaffiltext{1}{Physikalisches Institut, University of Bern, Sidlerstrasse 5, CH-3012 Bern, Switzerland. email: Yann.Alibert@phim.unibe.ch, Olivier.Mousis@obs-besancon.fr, Willy.Benz@phim.unibe.ch.}


\clearpage

\begin{abstract}

Using the clathrate hydrates trapping theory, we discuss the enrichments in volatiles
 in the atmosphere of Jupiter measured by the \textit{Galileo} probe in
the framework of new extended core-accretion planet formation models including
migration and disk evolution. We construct a self-consistent model in which the
volatile content of planetesimals accreted during the formation of Jupiter is
calculated from the thermodynamical evolution of the disk. Assuming CO$_2$:CO:CH$_4$
= 30:10:1 (ratios compatible with ISM measurements), we show that we can explain
the enrichments in volatiles in a way compatible with the recent constraints set from
internal structure modeling on the total amount of heavy elements present in the
planet.


\end{abstract}

\clearpage

\section{Introduction}

The formation and structure of giant planets, in particular those of the solar
system for which numerous observational constraints exist, has been the subject
of many studies during the recent years. In the present paper, we show that the
core accretion scenario (Pollack et al. 1996, hereafter referred to as P96) extended
to include migration and disk evolution can actually lead to the formation of a
Jupiter-like planet which accounts for both the constraints set by internal
structure models of this planet and the measurements of the abundances of volatiles
in its atmosphere.

Detailed internal structure models of Jupiter matching many observational constraints
(mass, radius, surface temperature, gravitational moments, etc.) have been constructed
recently (Guillot at al. 2004, hereafter referred to as G04). From this modeling, the
mass of the core of the planet and the total amount of heavy elements present can be
constrained. G04 found that this maximum amount of heavy elements present in Jupiter
is of the order of $\sim 42 \mearth$, whereas the mass of the core can vary anywhere
from $0$ to $13 \mearth$, the uncertainties being essentially due to our poor knowledge
of the equation of state of hydrogen. Note that the inferred present day core mass can
differ significantly from the one at the end of the formation process depending upon
the extent of core erosion (G04). The total amount of heavy elements present, on
the other hand, is not affected by this process.

Abundances of volatile species in Jupiter's atmosphere have been measured using the
mass spectrometer on-board the {\it Galileo} probe (Atreya et al. 1999, Mahaffy et al.
2000). These  measurements reveal that the giant planet's atmosphere is enriched by a
factor of $\sim 3$ in Ar, Kr, Xe, C, N and S  compared to the solar abundances of 
Anders \& Grevesse (1989, hereafter AG89). In order to
explain these enrichments, Owen et al. (1999) suggested that these volatile species
were trapped in planetesimals made of amorphous ice which were accreted by Jupiter.
Given the low temperature required for the existence of amorphous ice, this suggestion
implies one of the following three possibilities: 1) Jupiter was formed in sufficiently
cold regions (about 30 AU) and subsequently migrated inwards to its present location, or
2) the solar nebula was substantially cooler at 5 AU than currently imagined, or 3)
Jupiter accreted a substantial amount of ices originating from distant regions.

Another explanation has been proposed by Gautier et al. (2001a,b) and Hersant et al
(2004) who suggested that volatile species can also be trapped by H$_2$O ice at much higher
temperature in forms of clathrate hydrates or hydrates. Hence, the measured enrichments
could also be explained by the accretion of icy planetesimals originating from the
local, warmer region of the nebula. However, the large amount of water ice required to
trap the volatile species in these models implies a total heavy element content in
Jupiter incompatible with the constraints derived by G04. Finally, while these models
are based on the P96 calculations, they are not fully consistent with these calculations
for a number of reasons. First, they assume that planetesimals are captured by Jupiter
during the late stages of the formation process, while in the models of P96, $\sim 10
\mearth$ of heavy elements are accreted during the early phases. Second, the
thermodynamical conditions of the disk
evolution models used in Gautier et al. (2001a,b) and Hersant et al. (2004)
differ from the ones used as outer boundary conditions for
the planet formation calculations of P96.

The standard giant planets formation model by P96 has been recently extended by Alibert
et al. (2004,2005a) who included migration of the growing
planet and disk evolution. The authors show that these two phenomena can have significant
consequences on the formation process. This motivated us to recalculate the enrichments
in volatiles in Jupiter in a way fully consistent with these new formation models. For
this, we use the calculations by Alibert et al. (2005b,  hereafter referred to as A05)
to compute the thermodynamical properties of the 
disk, which allow the composition of the ices to be determined, and the time dependent accretion
rate of planetesimal during the growth of the planet. As we will show, this fully 
self-consistent approach can explain the measured enrichments while requiring a total 
abundance of heavy elements in agreement with the constraints derived by G04.

\section{Theoretical models}
\label{model}

\subsection{Formation of Jupiter}

The model of the formation of Jupiter we consider in this work is the one presented in A05 
which is based on the extended core-accretion formation approach described by Alibert et al. 
(2004,2005a). In this model, the giant planet forms from an embryo originally located at $\sim  
9$ AU \footnote{The precise starting location of the embryo depends upon the assumed 
migration rate and can range from 9 to 14 AU, see A05.} which migrates inwards and stops 
at the current position of Jupiter at the time the disk disappears.

This model takes into account the evolution of the protoplanetary disk in the framework of 
the $\alpha$ formalism (Shakura \& Sunyaev 1973). The thermodynamical properties of the 
disk as function of position and time ($T,P$ and $\Sigma$) are calculated by solving the 
vertical structure equations using the method presented in Alibert et al. (2005a). These 
quantities are used to determine the composition of the ices incorporated in the planetesimals 
(see next Section).

\subsection{Composition of planetesimals in Jupiter's feeding zone}
\label{calcul_chimie}

We assume that volatiles have been trapped during the cooling of the nebula in planetesimals
either in form of pure condensates or, following Gautier et al. (2001a,b), in form of 
hydrates or clathrate hydrates. Once condensed, ices are assumed to decouple from the gas, 
to be incorporated into growing planetesimals which may subsequently be accreted by the 
forming Jupiter.

\begin{figure}
\centerline{\includegraphics[width=6cm,angle=-90]{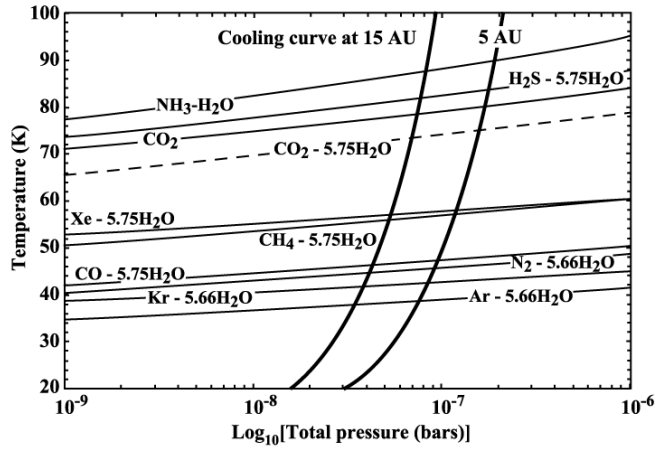}}
\caption{
Stability  curves of the condensates considered in the present work, and evolutionary tracks of the nebula at 5 and 15 AU.
Abundances of various elements are solar, with CO$_2$:CO:CH$_4$ = 30:10:1 and N$_2$:NH$_3$ = 1 in vapor phase.
The condensation curve of CO$_2$ pure condensate (solid line) is shown together with that of the corresponding
clathrate hydrate (dashed line). Species remain in the vapor phase as long as they stay in the domains located above the curves of stability.}
\label{cooling}
\end{figure}

Figure \ref{cooling} shows cooling curves of the nebula at 5 and 15 AU derived from the 
A05 model, as well as the condensation curves for the various ices considered in this work. 
The stability curves of clathrate hydrates are derived from Lunine \& Stevenson (1985) 
whereas for the pure CO$_2$ condensate a fit to the existing experimental data (Lide 1999, 
pp 6-59) is used. This figure provides the condensation sequence of the different volatiles 
initially existing in vapor phase inside Jupiter's feeding zone and the thermodynamical 
conditions ($T, P$ and $\Sigma$) at which the different ices are formed correspond to the 
intersection between the cooling curve and the stability curve of the different condensates.

In this work, we assume that the amount of water available along the migration pathway of 
the forming Jupiter is at least large enough to trap all the volatile species. With our 
assumption of decoupling upon condensation, this implies a minimum abundance of water 
relative to $H_2$ necessary of

\begin{equation}
x_{H_2O} = \sum_{\rm element~i} \gamma_i~x_i~\frac{\Sigma(R,P_i,T_i)}{\Sigma(R,P_{H_2O},T_{H_2O})},
\label{water}
\end{equation}
where $x_{i}$ is the molar mixing ratio of the volatile $i$ in vapor phase with respect to H$_2$
in the solar nebula, $\gamma_i$ is the required number of water molecules to form the
corresponding hydrate or clathrate hydrate (see Lunine \& Stevenson 1985; $\gamma_i = 0$ in
the case of pure condensate). $\Sigma(R,P_i,T_i)$ is the surface density of the solar
nebula at the distance $R$ from the Sun, at the intersection of the cooling curve and
the condensation curve of the species $i$.

Finally, for a given abundance of water, we  estimate $Y_i$, the volatile $i$ to water mass ratio
in icy planetesimals, by the relation given by Mousis \& Gautier (2004):
\begin{equation}
Y_i = \frac{X_i}{X_{H_2O}} \frac{\Sigma(R,P_i,T_i)}{\Sigma(R,P_{H_2O},T_{H_2O})},
\label{frac}
\end{equation}
where $X_i$ and $X_{H_2O}$ are respectively the mass mixing ratio of the volatile $i$ and
of H$_2$O with respect to H$_2$, in the vapor phase of the solar nebula.

\section{Results}
\label{results}

\subsection{Initial ratios of CO$_2$:CO:CH$_4$ and N$_2$:NH$_3$}
\label{init}

In order to be able to compute the composition of the planetesimals, a number of assumptions 
have to be made. In the present paper, we assume that in the solar nebula gas phase the
elements are in solar abundance (AG89, see Table \ref{anders}). Note that we also assume that
the amount of water ice is sufficient to trap all volatiles. As we shall see in Section
\ref{enrichments}, this corresponds to 
an abundance of water relative
to $H_2$ in vapor phase in excess of the value derived from solar abundances (see Table \ref{anders}).
Sedimentation and 
drift due to gas drag on the icy grains could explain this overabundance of water (Supulver
and Lin 2000). Furthermore, we assume that O,C and N exist only under the form of H$_2$O, 
CO$_2$, CO, CH$_4$, N$_2$ and NH$_3$. S is present only in the form of H$_2$S and other 
sulfur components (see next Section). 

Starting with ISM initial values, Mousis et al. (2002) have calculated the evolution 
of the CO:CH$_4$ and N$_2$:NH$_3$ ratios in vapor phase in the solar nebula by 
taking into account turbulent diffusion and kinetics of chemical conversions. They 
demonstrated that these ratios remain quasi-identical to the ISM values. Therefore,
we adopt CO:CH$_4$~=~10 in the vapor phase of the solar nebula, a value consistent 
with the ISM measurements of Allamandola et al.~(1999). Moreover, observations of 
ISM ices reveal the presence of large amount of CO$_2$ (see Gibb et al. 2004 for 
a review). Hence, CO$_2$ should be initially present in the vapor phase of the solar 
nebula, with CO$_2$:CO = 1 to 4, a range of values that covers the ISM measurements 
(Gibb et al. 2004).

Following the conclusions of Mousis et al. (2002), we also adopt for N$_2$:NH$_3$ 
a ratio close to the ISM value. However, even though current chemical models of the 
solar nebula predict that N$_2$ is more abundant than NH$_3$, no observations have
been published that would constrain this ratio further. In fact, some local processes 
may accelerate the conversion of N$_2$ to NH$_3$ in the solar nebula. This is for 
example the case if there exist some Fe grains that may have a catalytic effect on 
this conversion (Fegley 2000). Given these uncertainties, we compute the expected abundances 
of volatiles trapped in icy planetesimals using N$_2$:NH$_3$ = 1 and 10.

Finally, it is important to note that since the evolution of the nebula proceeds from 
high to low temperatures, CO$_2$ crystalizes as a pure condensate prior to be trapped 
by water to form a clathrate hydrate. Therefore, we assume that solid CO$_2$ is the only 
existing condensed form of CO$_2$ present in the solar nebula.

With these assumptions, Eq. \ref{water} and \ref{frac} together with the thermodynamical
evolution of the disk from A05 allow us to calculate the mass ratio of trapped volatiles 
to water, in the icy planetesimals accreted by the growing Jupiter (see Table 
\ref{planetesimaux}).

\subsection{Enrichments in volatiles in Jupiter}
\label{enrichments}

Using the results presented in Table \ref{planetesimaux}, we can calculate the enrichments 
in volatiles in Jupiter. Since the composition of planetesimals does 
not vary significantly along the migration path of Jupiter (between 15 AU to 5 AU, 
see Table \ref{planetesimaux}), we assume that all the ices of accreted planetesimals have 
an identical composition.  
Furthermore, in our definition of the ices/rocks ratio (I/R), we include all the types of ices
(H$_2$O, CO$_2$, hydrates and clathrate hydrates).

For the sake of comparison, we adopt first the same initial ratios for C and N 
species in vapor phase as in Hersant et al. (2004), namely CO$_2$:CO:CH$_4$~=~0:10:1, 
and N$_2$:NH$_3$~=~10. With these values, we reproduce their fit to the observed 
enrichments in volatiles. However, in order to obtain these values, $36.3 \mearth$ of 
water and $9.7 \mearth$ of other volatiles have to be accreted. Regardless of the assumed
value for the planetesimals I/R ratio, this yields a total amount
of heavy elements significantly exceeding the upper limit derived by G04. We note,
as did Hersant et al. (2004), that the calculated enrichment of sulfur is significantly 
higher than the observed value. This problem stems from the assumption that S was 
only present under the form of H$_2$S in the solar nebula. However, owing to the 
production of sulfur compounds from H$_2$S at high temperature and pressure early 
in the nebula, the abundance of  H$_2$S  may be subsolar by the time planetesimals 
form (Hersant et al.  2004).

If we now assume CO$_2$:CO:CH$_4$~=~30:10:1 and N$_2$:NH$_3$~=~1 (our nominal model)
in vapor phase in the solar nebula, we also reproduce the enrichments in volatiles (see 
Table \ref{enrichissements}). However, the minimum mass of accreted ices is found to 
be $25.4 \mearth$, including at least $11.3 \mearth$ of water and $14.1 \mearth$ of other 
volatiles. The resulting total content in heavy elements in Jupiter is therefore compatible
with G04, provided the I/R ratio of accreted planetesimals is greater than $\sim 1.5$.
This also translates into a minimum O/H value in Jupiter's envelope equal to  $\sim 4.2 \times 10^{-3}$ (ie
$\sim 5$ times the solar value of AG89) 
or about half the value obtained by Hersant et al. (2004). Finally,
we note that the amount of water needed corresponds, if vaporized, to an abundance 
relative to H$_2$ of at least  $\sim 9.5 \times 10^{-4}$, or $\sim 2$ times its 
gas-phase abundance in the solar nebula 
(see Table \ref{anders}). Again, 
we explain this difference by sedimentation and drift effects of icy grains (Supulver
and Lin 2000).

We have  estimated the sensitivity of our results to changes in the initial CO$_2$:CO:CH$_4$
and N$_2$:NH$_3$ ratios. Keeping  N$_2$:NH$_3$ = 1, we find that it is possible to 
fit the measured enrichments while remaining compatible with G04 for CO$_2$:CO = 1,2,3 
or 4 (see Table \ref{ices}). However, if N$_2$:NH$_3$ is increased to 10, only 
CO$_2$:CO=1 gives surface abundances compatible with the measured enrichments and
a total amount of heavy elements compatible with G04 but only provided the mean I/R
ratio of planetesimals is at least equal to $\sim$ 4.4 (see Table \ref{ices}).

Recently, the solar abundances have been subject to some revisions (Allende Prieto et al. 2001, 2002; see  Lodders 2003 (L03 in the following)
 for a review).
With these new values,
the enrichments of C, N, S, Ar, Kr, and Xe are now $3.7 \pm 1.0$, $5.5 \pm 2.4$, $2.2 \pm 0.3$,
$2.2 \pm 0.8$, $2.0 \pm 0.7$, and $2.0 \pm 0.5$,  taking into account the uncertainties
on the solar abundances of L03. 
These enrichments can still be accounted for in our nominal model using
the new solar abundances, except for N which is found to be enriched only by
a factor 2.5 over the value given gy L03.
However, note that N/H in Jupiter has been recently revised down (Wong et al. 2004), the
resulting enrichment of N (relative to the values of L03) being 
$3.9 \pm 1.9$. In that case, our nominal model is compatible with all 
the abundances of the afore-mentionned elements in Jupiter.
Finally, this requires the accretion of $20.6 \mearth$ of ices, and translates into a
minimum  O/H in Jupiter's envelope equal to $\sim 3.3 \times 10^{-3}$.
In order to be compatible with G04, the I/R ratio of accreted
planetesimal must be greater than $\sim 1$.

\section{Summary and discussion}
\label{conclusion}

In this Letter, we have presented calculations of the enrichments in volatile species 
in Jupiter's atmosphere using the latest giant planet formation models of A05 which 
include migration and disk evolution. The computed enrichments are therefore fully consistent 
with the thermodynamical evolution of the disk and with the growth of the giant planet.

In our nominal model (CO$_2$:CO:CH$_4$ = 30:10:1, and N$_2$:NH$_3$ =1), the enrichments 
in volatiles observed in Jupiter's atmosphere result from the accretion of 25.4
$\mearth$ of icy planetesimals. Assuming that these planetesimals contain at least $\sim 
60 \%$ of ices (I/R $\geq 1.5$), the total amount of heavy elements accreted is compatible 
with the determinations of G04. This consistency between surface abundance measurements 
and total content in heavy elements remains valid for a range of CO$_2$:CO and  N$_2$:NH$_3$ 
ratios albeit requiring in some cases stronger constraints on the minimum I/R ratio of 
accreted planetesimals (from $\sim 1.4$ to $\sim 4.4$). We note that if we ignore the presence 
of pure CO$_2$ ice but assume that C is trapped in form of CO and CH$_4$ only, we were 
capable to match the measured enrichments while remaining compatible with G04 only if we 
assumed the accretion of pure ice planetesimals (I/R = $\infty$).

Furthermore, our nominal model predicts that the O/H ratio in the atmosphere of Jupiter 
must be at least $\sim 4.2 \times 10^{-3}$ which is half the one obtained 
by Hersant et al. (2004). This ratio can possibly be measured by future space missions.
Interestingly, 
the minimum I/R ratio ($\sim 1.5$) required to satisfy the upper limit 
derived by G04 happens to be consistent with the estimated ices content of the Galilean icy satellites 
($50$ to $60 \%$, see Sohl et al. 2002). Our results are then compatible with the 
formation of regular icy satellites from the accretion of planetesimals originating from 
the nebula without having been vaporized inside the subdisk (Mousis \& Gautier 2004, 
Mousis et al. 2005).

We conclude that the enrichments in volatiles in Jupiter's atmosphere  can then be 
explained in a way compatible with both the internal structure models of the giant 
planet, and the average composition of its regular icy satellites. Furthermore, if one 
assumes a value of I/R in planetesimals of the order of $\sim 1.5$, the enrichments 
in volatiles are also compatible with the high mass model of G04.
 This model is however
characterized by a low core mass while the model of A05 predicts a core mass of  
$\sim 6 \mearth$. Thus, we must assume that the core has been eroded over time (see G04 
for details) and that the volatiles which have not been released during the accretion phase
itself have been released in the atmosphere during this process.

\acknowledgements 
This work was supported in part by the Swiss National Science Foundation.  O.M. was 
supported by an ESA external fellowship.

{}

\clearpage

\begin{table}[h] 
   \caption[]{Gas-phase abundances (molar mixing ratio with respect to H$_2$) of major species in the solar nebula (from AG89) 
for CO$_2$:CO:CH$_4$~=~30:10:1 and N$_2$:NH$_3$~=~1 (our nominal 
model).} 
\label{anders}
  \begin{center} 
       \begin{tabular}[]{lcclc} 
            \hline 
            \hline 
            \noalign{\smallskip} 
             Species $i$ & $x_i$ &  &  Species $i$ & $x_i$ \\ 
            \noalign{\smallskip} 
             \hline 
             \noalign{\smallskip} 
             O &   $1.71 \times 10^{-3}$ &  & N$_2$  &   $7.47 \times 10^{-5}$ \\ 
             C &  $7.26 \times 10^{-4}$ &  & NH$_3$  &  $7.47 \times 10^{-5}$ \\ 
             N &  $2.24 \times 10^{-4}$ &  & S & $3.24 \times 10^{-5}$ \\ 
             H$_2$O  & $4.86 \times 10^{-4}$ &  & Ar & $7.26 \times 10^{-6}$ \\ 
             CO$_2$  & $5.31 \times 10^{-4}$ &  & Kr & $3.39 \times 10^{-9}$ \\ 
             CO  & $1.77 \times 10^{-4}$ &  & Xe  & $3.39 \times 10^{-10}$ \\ 
             CH$_4$  & $1.77 \times 10^{-5}$ \\ 
 
            \hline 
         \end{tabular} 
         \end{center} 
         \end{table} 

\clearpage

\begin{table}[h]
         {\setlength{\tabcolsep}{0.33cm}
         \begin{center}
         \begin{tabular}[]{lcc}
            \tableline
            \tableline
               &  5 AU & 15 AU  \\
             \tableline
             \noalign{\smallskip}
             CO$_2$:H$_2$O  &  $9.43 \times 10^{-1}$ &  $9.28 \times 10^{-1}$  \\
             CO:H$_2$O  &  $1.47 \times 10^{-1}$ &  $1.48 \times 10^{-1}$  \\
             CH$_4$:H$_2$O     &  $9.44 \times 10^{-3}$  &  $9.39 \times 10^{-3}$  \\
             N$_2$:H$_2$O      &  $6.07 \times 10^{-2}$  &  $6.11 \times 10^{-2}$   \\
             NH$_3$:H$_2$O    &  $5.45 \times 10^{-2}$   &  $5.34 \times 10^{-2}$  \\
             H$_2$S:H$_2$O     &  $3.16 \times 10^{-2}$  &  $3.11 \times 10^{-2}$  \\
             Ar:H$_2$O    &  $6.83 \times 10^{-3}$  &  $6.95 \times 10^{-3}$  \\
             Kr:H$_2$O    &  $8.01 \times 10^{-6}$  &  $8.10 \times 10^{-6}$  \\
             Xe:H$_2$O    &  $1.52 \times 10^{-6}$  &  $1.51 \times 10^{-6}$  \\
            \tableline
         \end{tabular}
         \caption[]{
Calculations of the ratios of trapped masses of volatiles to the mass of H$_2$O ice ($Y_i$ in 
 the text) in planetesimals
 formed at 5 and 15 AU in the solar nebula. Gas-phase abundance of H$_2$O is equal to $9.5 \times 10^{-4}$ ($\sim 2$ times the value
quoted in Table \ref{anders}, see text), and  gas-phase abundances of elements are assumed to be solar 
(AG89) with CO$_2$:CO:CH$_4$~=~30:10:1 and with N$_2$:NH$_3$~=~1 in vapor phase in
 the solar nebula. The abundance of H$_2$S is subsolar (see text).
}
\label{planetesimaux}
\end{center}}
\end{table}

\clearpage

\begin{table}%
         {\setlength{\tabcolsep}{0.3cm}
         \begin{tabular}{lcc}
            \tableline
            \tableline
            \noalign{\smallskip}
              Species &  Observed & Calculated   \\
             \noalign{\smallskip}
             \tableline
             \noalign{\smallskip}
             Ar  &  $2.5 \pm 0.5$$^{~a}$  &  2.15   \\
             Kr  &  $2.7 \pm 0.5$$^{~a}$  &  2.30    \\
             Xe &  $2.6 \pm 0.5$$^{~a}$ &   2.80   \\
             C   &  $2.9 \pm 0.5$$^{~b}$ &  3.10  \\
             N   &  $3.6 \pm 0.8$$^{~b}$ &  2.80   \\
             S   &  $2.5 \pm 0.15$$^{~b}$ & 2.35  (3.40 without correction)\\
            \hline
         \end{tabular}}
         \caption{Observed enrichments in volatiles in Jupiter,
and calculated enrichments in volatiles in the nominal model (CO$_2$:CO:CH$_4$ =30:10:1 and N$_2$:NH$_3$ = 1).
The observed values are taken from
Mahaffy et al. (2000) (a) and Atreya et al. (1999) (b), and are relative to solar abundances of AG89.
The calculated abundance of sulfur (3.4 without correction) is revised down to fit the
observed value (H$_2$S/H$_2 = 0.7 \times ($S/H$_2)_\odot$; see text).
}
\label{enrichissements}
\end{table}

\clearpage

\begin{table}%
         {\setlength{\tabcolsep}{0.3cm}
         \begin{tabular}{lccccc}
            \tableline
            \tableline
            \noalign{\smallskip}
              & \multicolumn{4}{c}{N$_2$:NH$_3$ = 1} & N$_2$:NH$_3$ = 10  \\
            \tableline
              & CO$_2$:CO = 4 & CO$_2$:CO = 3 & CO$_2$:CO = 2 & CO$_2$:CO = 1  & CO$_2$:CO = 1   \\
             \noalign{\smallskip}
             \tableline
             \noalign{\smallskip}
             $M_{\rm water} / \mearth$ & 9.9  & 11.3 & 13.5 & 17.9 & 20.8 \\
             $M_{\rm ices} / \mearth$ &  24.4  & 25.4  & 26.9 & 30.0 & 34.2 \\
             I/R & 1.4 & 1.5 & 1.8 & 2.5  & 4.4 \\
             O/H & 4.7 & 5 & 5.3 & 6.1 & 7 \\
            \hline
         \end{tabular}}
         \caption{Minimum mass of accreted water ($M_{\rm water}$),
minimum mass of accreted ices ($M_{\rm ices}$), and minimum I/R ratio required to fit both the observed enrichments
in volatiles, and the constraints of G04. The last line gives the minimum O/H abundance (compared to solar value of AG89)
in the atmosphere of Jupiter.}
\label{ices}
\end{table}

\clearpage

\end{document}